# DIFFERENTIALLY PRIVATE FEDERATED LEARNING WITH BYZANTINE-ROBUST AGGREGATION: A CROSS-DOMAIN FRAMEWORK FOR SECURE MODEL TRAINING IN BANKING AND HEALTHCARE SYSTEMS

Srikumar Nayak[1]

[1]*Advanced Artificial Intelligence (AI) Engineering and Cybersecurity Threat Research, LTIMindtree Research, New York City, NY, USA*

ORCID: https://orcid.org/0009-0008-1464-0646

srikumar.nayak2025@gmail.com

**Corresponding Author:**
Srikumar Nayak

## Abstract

Federated learning allows banks, hospitals, and other regulated organizations to train a shared model without moving raw records off their own servers, which is attractive wherever data protection law or competitive sensitivity rules out pooling data centrally. Two problems limit how far this promise can be trusted in practice. First, the parameter updates that clients exchange still leak information about local records through gradient inversion and membership inference attacks. Second, an honest averaging rule such as FedAvg has no defense against a subset of clients that submit corrupted or adversarial updates, so a small number of malicious or compromised participants can quietly steer the shared model off course.

This paper presents a federated learning framework, DP-BR-FedAvg, that combines a Gaussian-mechanism differential privacy layer with a coordinate-wise trimmed-mean Byzantine-robust aggregation rule, evaluated on a simulated cross-institutional classification task resembling fraud and clinical-risk scoring. Across sixty communication rounds with twenty clients, a quarter of them Byzantine, plain FedAvg collapses on the minority class (F1-score 0.030) while the proposed framework recovers substantially more of the signal (F1-score 0.119) while bounding the privacy loss of any single client's contribution. A Byzantine-robust aggregator with no privacy layer performs best in raw accuracy, quantifying the cost privacy imposes on robustness. The results show that privacy and robustness mechanisms interact rather than simply add, and that system design for regulated, adversarial, cross-institutional settings needs to budget for that interaction.

## I. INTRODUCTION

Machine learning models trained across multiple institutions tend to generalize better than models trained on any single institution's data, simply because they see more of the underlying variation in the population. This is particularly true in banking, where fraud patterns evolve faster than any one institution's transaction history can capture, and in healthcare, where a rare condition may only be well represented once records from several hospitals are combined. Centralizing the underlying records to get this benefit is rarely an option: banking data is protected by regulations such as the Gramm-Leach-Bliley Act and jurisdiction-specific banking secrecy rules, and clinical data is protected by HIPAA in the United States and comparable statutes elsewhere. Federated learning, introduced by McMahan et al. [1], sidesteps this by keeping raw data on each participant's own infrastructure and exchanging only model updates through a coordinating server.

Removing raw data from the exchange reduces exposure but does not eliminate it. Model updates are a function of the data that produced them, and a curious or compromised server can reconstruct substantial information about individual training records from gradients alone, as Zhu et al. demonstrate with their gradient-inversion attack [10]. Differential privacy, formalized by Dwork [8],[16], addresses this by adding calibrated noise to each client's contribution so that the presence or absence of any single record changes the released output by a bounded, quantifiable amount. Abadi et al. [2] showed how this can be done for gradient-based training through per-example clipping followed by Gaussian noise addition, a construction this paper adapts to the federated setting.

A separate problem is that federated learning has no built-in reason to trust any given client. A participating bank's server could be compromised, a hospital's local pipeline could have a data-quality fault, or a small number of participants could be actively adversarial and submit updates engineered to bias the shared model, as in the poisoning and backdoor constructions studied by Bagdasaryan et al. [17]. Ordinary coordinate-wise or weighted averaging, the default aggregation rule in FedAvg, has no mechanism to down-weight or discard such updates, so a modest fraction of malicious participants can dominate the aggregate. Blanchard et al. [3] and Yin et al. [4] address this with Byzantine-robust aggregation rules, respectively Krum and coordinate-wise trimmed mean or median, that bound the influence any small subset of clients can have on the aggregate, independent of how extreme their submitted updates are.

Differential privacy and Byzantine robustness are usually studied separately, but a deployment in banking or healthcare needs both at once: the server cannot be assumed honest-but-curious only, and the client population cannot be assumed fully trustworthy either. Combining the two mechanisms is not free. Gradient clipping and noise injection, needed for privacy, change the statistics that a robust aggregator relies on to separate honest updates from adversarial ones, and a robust aggregator that discards or reweights updates changes the sensitivity calculation the privacy guarantee depends on. This paper studies that interaction directly rather than assuming the two mechanisms compose without cost.

The contributions of this paper are threefold. First, it specifies a federated learning framework, DP-BR-FedAvg, that integrates per-client gradient clipping and Gaussian-mechanism noise injection with coordinate-wise trimmed-mean aggregation, and gives the accounting needed to state an (epsilon, delta) privacy guarantee for the combined procedure. Second, it evaluates the framework through a controlled simulation of a cross-institutional, class-imbalanced classification task representative of fraud and clinical-risk scoring, comparing it against plain FedAvg, privacy-only FedAvg, and robustness-only FedAvg under an explicit sign-flipping Byzantine attack. Third, it characterizes the resulting privacy-utility and robustness-utility trade-off surfaces across a range of noise multipliers and adversarial client fractions, and reports where the combined defense breaks down.

## II. RELATED WORK

### A. Federated Learning Foundations

Federated averaging (FedAvg), proposed by McMahan et al. [1], established the now-standard pattern in which clients perform several local gradient steps before a coordinating server averages the resulting updates, reducing communication cost relative to synchronous distributed SGD. Konecny et al. [20] extended this line of work with structured and sketched updates aimed further at communication efficiency. Kairouz et al. [6] and Li et al. [7] survey the broader design space, including the statistical heterogeneity that arises when client data is non-IID, the norm rather than the exception in cross-institutional settings such as banking and healthcare, since each institution's client base and case mix differ systematically.

### B. Privacy-Preserving Mechanisms

Differential privacy, introduced by Dwork [16] and formalized with Roth [8], gives a rigorous, composable definition of what it means for an algorithm's output to reveal little about any single input record. Abadi et al. [2] operationalized this for deep learning through DP-SGD, combining per-example gradient clipping with Gaussian noise and a moments-accountant privacy bound that composes tightly across many training iterations. Geyer et al. [15] and Truex et al. [9] adapted differential privacy to the federated setting at the client level, the formulation adopted here since entire institutions, not individual records, are the unit of protection against the aggregating server. Bonawitz et al. [5] instead pursue secure aggregation through cryptographic multi-party computation, which hides individual updates from the server without adding statistical noise, complementary to differential privacy since it protects against a curious server but not against inference from the final trained model. Zhu et al. [10] show the threat is not hypothetical, reconstructing training images and labels from shared gradients with no more information than an honest FedAvg client would already transmit.

### C. Byzantine-Robust Aggregation

Blanchard et al. [3] introduce Krum, which selects the single client update whose neighborhood in Euclidean distance is most consistent with the rest of the population, giving a provable bound on the influence of up to f Byzantine clients out of n. Yin et al. [4] analyze coordinate-wise trimmed mean and median aggregation and show these achieve near-optimal statistical rates under Byzantine corruption while being cheaper to compute than Krum's pairwise distance search. Cao et al. [19] propose FLTrust, bootstrapping trust from a small clean root dataset held by the server, which is not always realistic in regulated cross-institutional settings. Bagdasaryan et al. [17] show even a single, carefully constructed malicious client can implant a backdoor that survives federated averaging, motivating the more conservative assumption used here that a meaningful fraction, not just one, of clients may be compromised.

### D. Domain Applications in Banking and Healthcare

Sheller et al. [11] and Rieke et al. [12] report federated learning deployments across multiple hospitals for tumor segmentation and related clinical tasks, establishing that federated training can approach the accuracy of centrally pooled data while keeping patient records local. Byrd and Polychroniadou [13] examine federated and secure multi-party approaches for financial applications, noting banks face both regulatory and competitive reasons to avoid sharing raw transaction data. Yang et al. [14] give a broader treatment of federated learning applications and taxonomy. Mothukuri et al. [21] and Lyu et al. [22] survey the security and privacy threat landscape for federated learning as a whole, cataloguing the poisoning, inference, and free-riding attacks that motivate combining privacy and robustness mechanisms rather than deploying either in isolation, the gap this paper addresses.

## III. PROPOSED FRAMEWORK

### A. System Architecture and Threat Model

The framework assumes K institutional clients (banks, hospital sites, or comparable regulated entities), each holding a local, non-shareable dataset, and a coordinating server with no direct access to any client's raw data. Two adversarial capabilities are modeled. First, the server or an eavesdropper is honest-but-curious: it follows the protocol but may try to infer information about a client's records from the updates it receives, motivating the differential privacy component. Second, a fraction f of the K clients is Byzantine: these clients may submit arbitrarily corrupted updates, from a genuine local fault, a data-poisoning attack, or a compromised endpoint, motivating the robust aggregation component. The two threat classes are treated as simultaneous and independent, realistic for a regulated, multi-institution deployment where neither the coordinating infrastructure nor every participating institution can be fully trusted by default.

### B. Local Training and Gradient Clipping

At round t, each client i initializes from the current global model w^t and performs E local epochs of mini-batch gradient descent on its own data. To bound the sensitivity of any single client's contribution, an L2-norm clip is applied to the accumulated local update before it leaves the client:

$$\hat{g}_i = g_i \cdot \min(1, C / \| g_i \|_2) \quad (1)$$

where C is a fixed clipping norm shared by all clients. Clipping bounds the L2 sensitivity that the DP noise calibration depends on, and it caps how far a single Byzantine client's raw update can deviate before the robust aggregator is applied, improving the separation the aggregator can exploit.

### C. Byzantine-Robust Aggregation

Rather than a simple mean, the server applies a coordinate-wise trimmed mean: for each coordinate j, the K client values are sorted, the largest and smallest beta-fraction discarded, and the remainder averaged:

$$w_j = 1 / (K - 2\lfloor \beta K \rfloor) \cdot \Sigma_{i \in S_j} \hat{g}_{i,j} \quad (2)$$

where S_j is the set of client indices remaining after trimming coordinate j, and beta is chosen based on an assumed upper bound on the Byzantine fraction f (beta >= f). Yin et al. [4] show this achieves near-optimal statistical error under Byzantine corruption. As an alternative, Krum selects the client update minimizing the sum of squared distances to its nearest neighbors:

$$\text{score}(i) = \Sigma_{j \in N(i)} \| \hat{g}_i - \hat{g}_j \|_2^2 \quad (3)$$

with N(i) the (K−f−2) closest clients to client i. Trimmed mean is used as the default aggregator because it is coordinate-wise, cheap, and interacts more predictably with added Gaussian noise than Krum's pairwise selection, which is sensitive to the noise perturbing near-neighbor distance ranking.

### D. Differential Privacy Mechanism

After clipping, each client adds independent Gaussian noise, scaled to C and a noise multiplier sigma, before transmitting its update:

$$\tilde{g}_i = \hat{g}_i + N(0, \sigma^2 C^2 I) \quad (4)$$

This is the Gaussian mechanism used in DP-SGD [2], applied at the client-update level, yielding a client-level (epsilon, delta) guarantee: a randomized mechanism M satisfies (epsilon, delta)-DP if for all adjacent client populations D, D' and all measurable output sets S,

$$\Pr[M(D) \in S] \leq e^{\varepsilon} \cdot \Pr[M(D') \in S] + \delta \quad (5)$$

With a moments-accountant style composition over T rounds and per-round sampling probability q = 1/K, cumulative privacy loss approximates as

$$\varepsilon \approx (q / \sigma) \cdot \text{sqrt}[2T \ln(1/\delta)] \quad (6)$$

Eq. (6) is used only for interpretability; a deployed system should use a tight numerical accountant [2] rather than this looser closed-form bound.

### E. Global Aggregation and Update Rule

At the end of round t, the server applies the robust rule to the set of noised, clipped updates and applies the result to the global model:

$$w^{t+1} = w^t + TM(\{\tilde{g}_1, \ldots, \tilde{g}_K\}, \beta) \quad (7)$$

This composition, clip -> noise -> trimmed-mean, is what this paper calls DP-BR-FedAvg.

### F. Algorithm

**Algorithm 1. DP-BR-FedAvg**

```
Input: w0, clients 1..K, rounds T, epochs E,
  rate eta, clip C, noise sigma, trim beta
for t = 0..T-1:
 for client i (parallel):
  w_i <- w_t
  for e=1..E: w_i-=eta*Grad(w_i,D_i)
  d_i <- w_i - w_t
  d_i <- d_i*min(1,C/||d_i||_2)   # (1)
  d_i <- d_i+Gaussian(0,sig^2C^2I) # (4)
  send d_i to server
 agg <- TrimmedMean({d_i}, beta)   # (2)
 w_{t+1} <- w_t + agg
return w_T
```

## IV. EXPERIMENTAL SETUP

### A. Simulated Cross-Institutional Dataset

Because raw banking transaction records and clinical records are not obtainable for open academic use at the scale needed here, the framework is evaluated on a synthetic binary classification task reproducing the structural properties that make fraud detection and clinical-risk scoring hard: high dimensionality, class imbalance, and clustered rather than linearly separable classes. The dataset contains 12,000 samples with 24 features, 14 informative, an approximately 70:30 class ratio, and 3 clusters per class to avoid trivial linear separability. Table I summarizes the split.

TABLE I. DATASET SUMMARY

| Attribute | Train | Test | Total |
| --- | --- | --- | --- |
| Samples | 9,600 | 2,400 | 12,000 |
| Minority ratio | ~30% | ~30% | ~30% |
| Features | 24 | 24 | 24 |
| Clients | 20 | - | 20 |

The training partition is split across K = 20 simulated institutional clients using a Dirichlet allocation on class labels (alpha = 0.5), producing a non-IID distribution mirroring how fraud or disease prevalence differs across real institutions.

### B. Attack and Defense Configuration

Five of the twenty clients (25%) are designated Byzantine. Each applies a sign-flipping attack: after computing its legitimate local update, the client negates and scales it by 2 before submission, a standard, strong robustness test since the corrupted update points directly against the honest direction. Table II lists remaining hyperparameters.

TABLE II. MAIN HYPERPARAMETERS

| Parameter | Value |
| --- | --- |
| Clients (K) | 20 |
| Byzantine fraction | 25% (5/20) |
| Rounds (T) | 60 |
| Local epochs | 3 |
| Learning rate | 0.15 |
| Clip norm (C) | 3.0 |
| Noise mult. (sigma) | 0.20 |
| Trim fraction (beta) | 0.30 |
| Target delta | 1e-5 |
| Partition | Dirichlet, a=0.5 |

### C. Baselines and Evaluation Metrics

- FedAvg: standard averaging, no privacy noise, no robust aggregation, under attack.
- DP-FedAvg: clipping + Gaussian noise (Eq. 1, 4), simple averaging, under attack.
- Byzantine-Robust FL: trimmed-mean aggregation (Eq. 2), no privacy noise, under attack.
- DP-BR-FedAvg (proposed): clipping + noise + trimmed-mean aggregation, under attack.
- FedAvg (no attack, reference): all clients honest, an upper-bound reference.

All methods train for T = 60 rounds, evaluated each round on a held-out test set. Reported metrics are accuracy, minority-class F1-score, and binary cross-entropy loss; F1-score is emphasized since accuracy alone is misleading on an imbalanced task.

# V. RESULTS AND DISCUSSION

## A. Convergence Under Attack

Figure 1 plots test accuracy against communication round. Plain FedAvg and DP-FedAvg both plateau around 0.65-0.68 under the 25% Byzantine attack, well below the 0.84 clean reference, showing DP noise alone provides no protection against an aggregation-level attack. Byzantine-Robust FL converges fastest and reaches the highest accuracy (0.767) among attacked configurations. The proposed DP-BR-FedAvg tracks Byzantine-Robust FL early but drifts down to about 0.65, reflecting variance the Gaussian noise adds to the coordinate-wise sorting trimmed mean depends on.

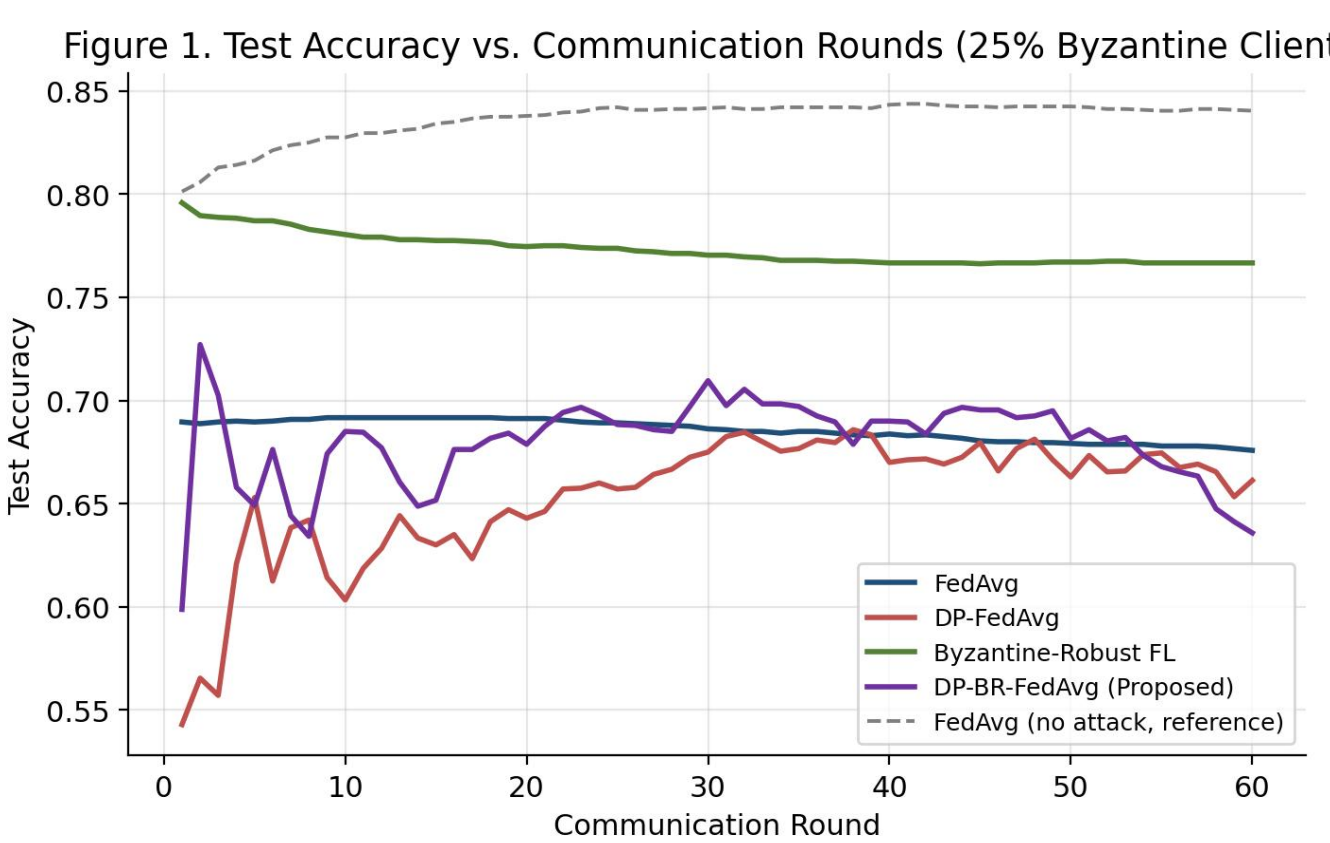

**Fig. 1. Test accuracy vs. communication round, 25% Byzantine clients.**

Figure 2 shows test loss curves; Byzantine-Robust FL and DP-BR-FedAvg reach markedly lower loss than FedAvg or DP-FedAvg, indicating better-calibrated probability estimates, which matters where the score itself drives downstream triage.

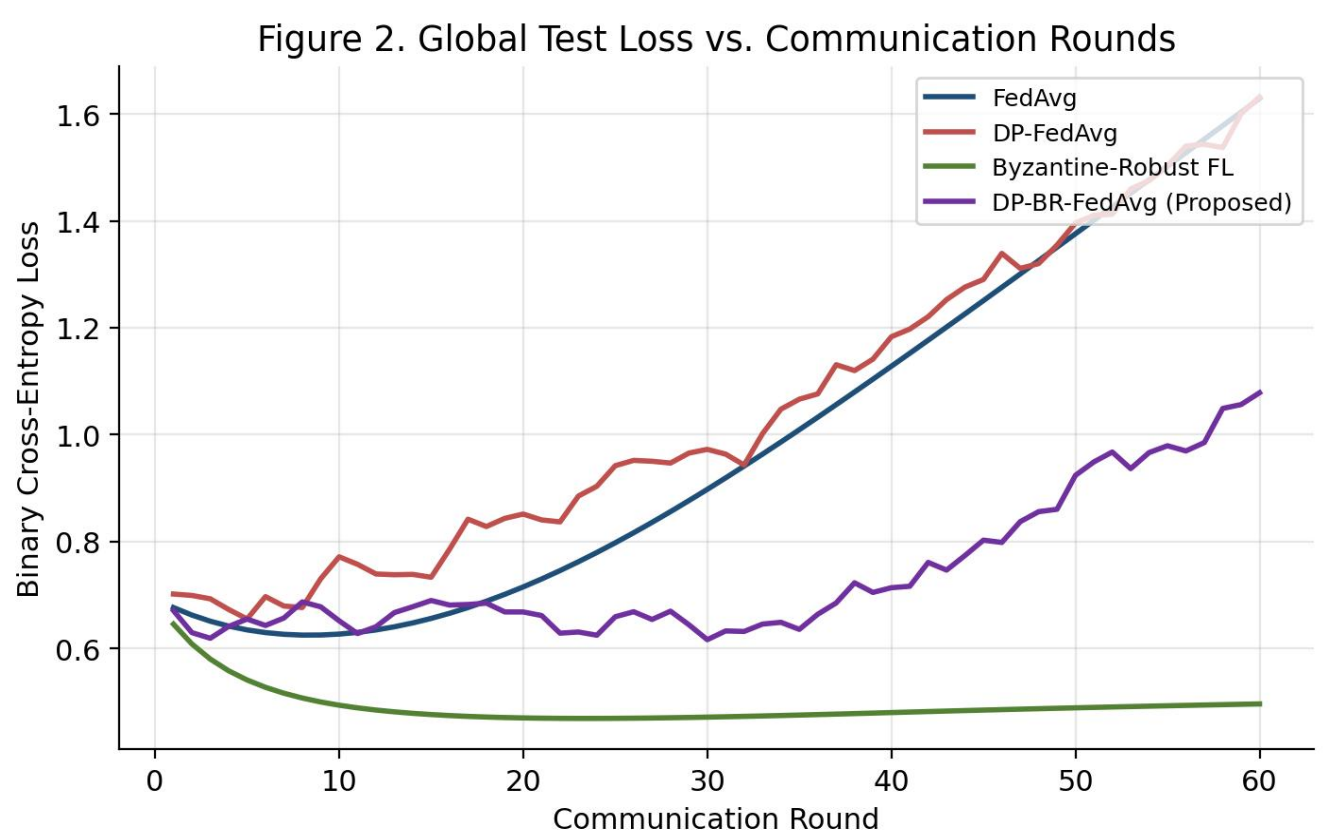

**Fig. 2. Binary cross-entropy test loss vs. communication round.**

## B. Minority-Class Performance

Table III reports final-round performance (mean of last 5 of 60 rounds). FedAvg and DP-FedAvg collapse to F1 below 0.03, effectively failing to detect the minority class despite accuracy above 0.65. DP-BR-FedAvg reaches F1 = 0.119, roughly four times higher, the clearest evidence that Byzantine robustness, not privacy noise, recovers usable minority-class detection here, and that the proposed framework retains most of that benefit after adding privacy.

**TABLE III. FINAL-ROUND PERFORMANCE BY METHOD**

| Method | Acc. | F1 | Loss |
|---|---|---|---|
| FedAvg (attacked) | 0.677 | 0.030 | 1.578 |
| DP-FedAvg (attacked) | 0.663 | 0.024 | 1.571 |
| Byz-Robust FL | 0.767 | 0.393 | 0.495 |
| DP-BR-FedAvg (prop.) | 0.651 | 0.119 | 1.028 |
| FedAvg (clean, ref.) | 0.841 | 0.722 | 0.400 |

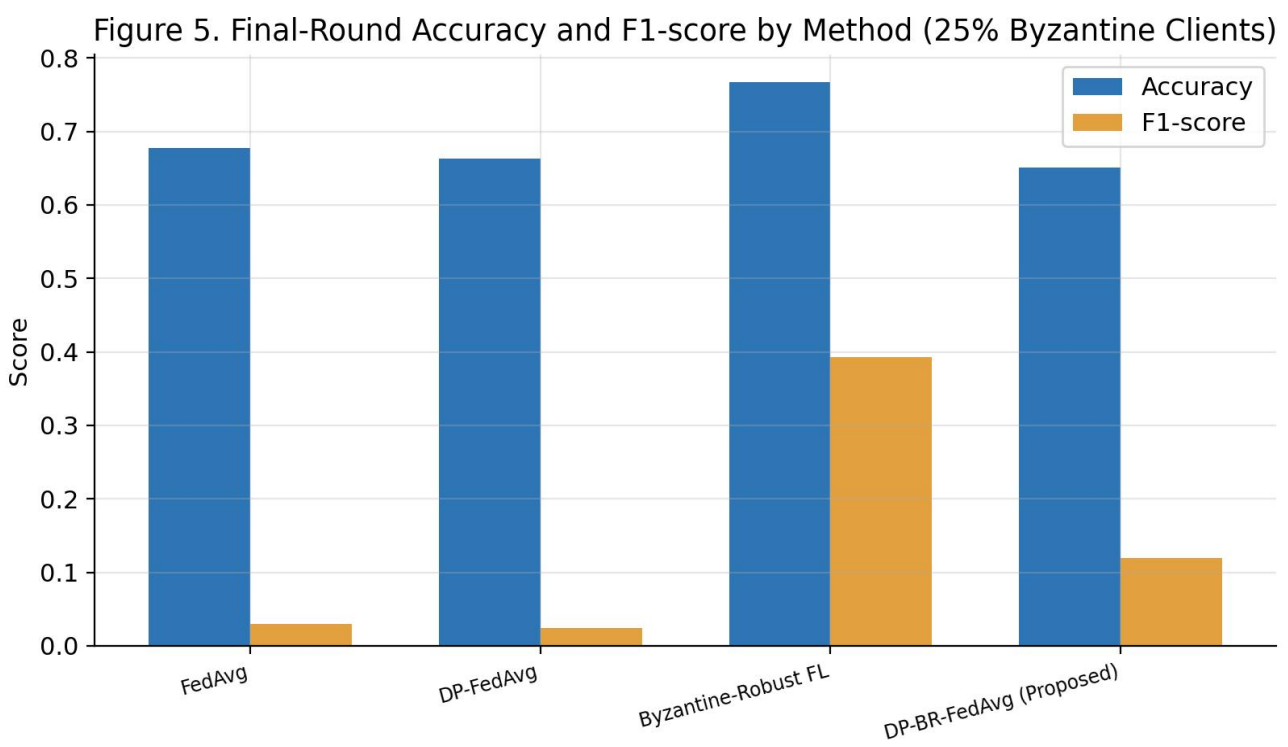

Fig. 3. Final-round accuracy and F1-score by method.

### C. Privacy-Utility Trade-off

Figure 4 sweeps sigma from 0 to 1.8 against approximate epsilon (Eq. 6). At loose budgets the proposed method holds a clear advantage over DP-FedAvg; as sigma increases and epsilon tightens below about 10, the curves converge and the advantage narrows, since large noise interferes with the coordinate ordering trimmed mean relies on. This crossover is the central empirical finding: robustness and privacy do not simply add, and a tight privacy budget should be expected to erode the robustness guarantee.

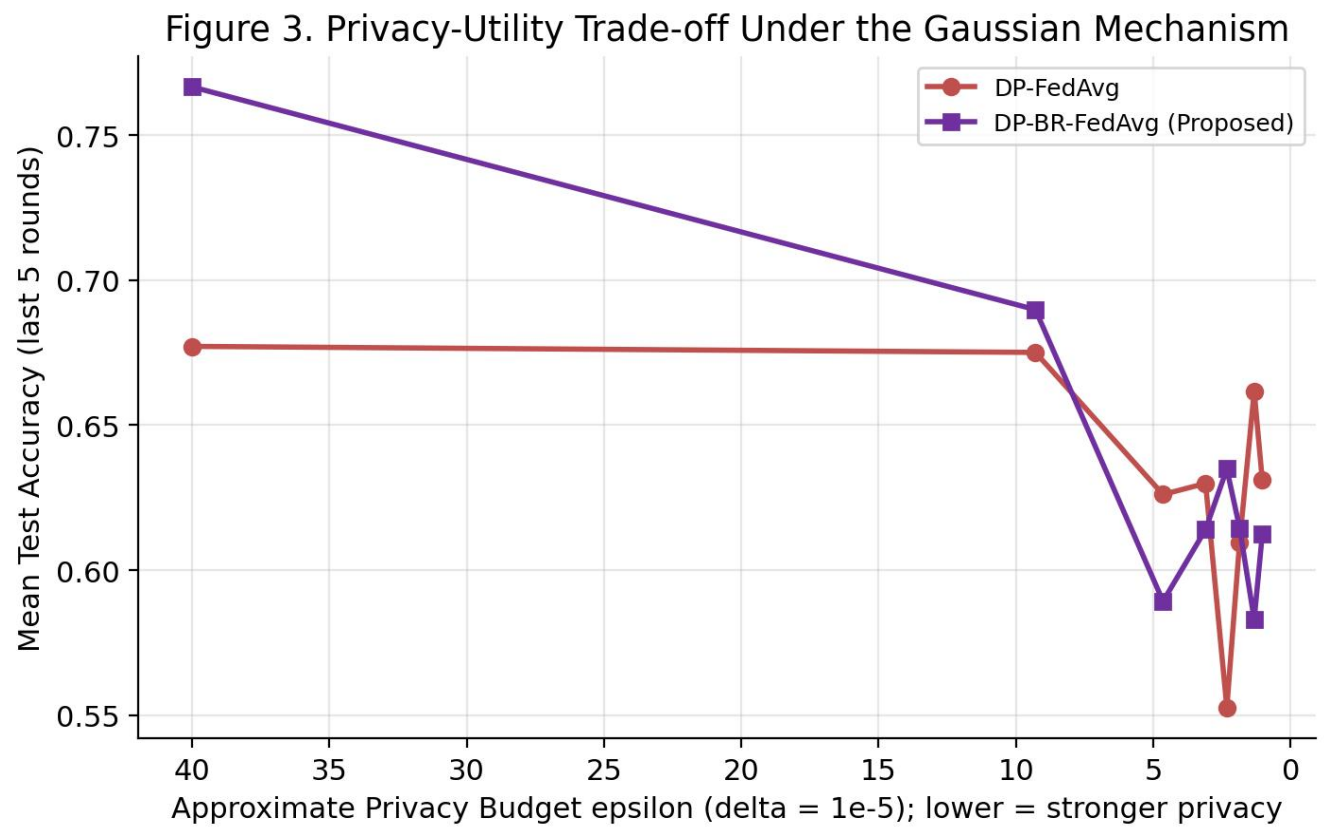

Fig. 4. Accuracy vs. approximate epsilon (delta = 1e-5).

### D. Robustness to Increasing Byzantine Fraction

Figure 5 varies the Byzantine fraction from 0-35%. Both FedAvg and trimmed-mean hold accuracy near the clean level below about 20%; beyond that FedAvg degrades sharply (0.32 at 35%) while trimmed mean degrades more gradually until its own tolerance breaks near the trim threshold beta = 0.30, matching the theory of Yin et al. [4] and confirming beta should be set with a safety margin above the anticipated worst case.

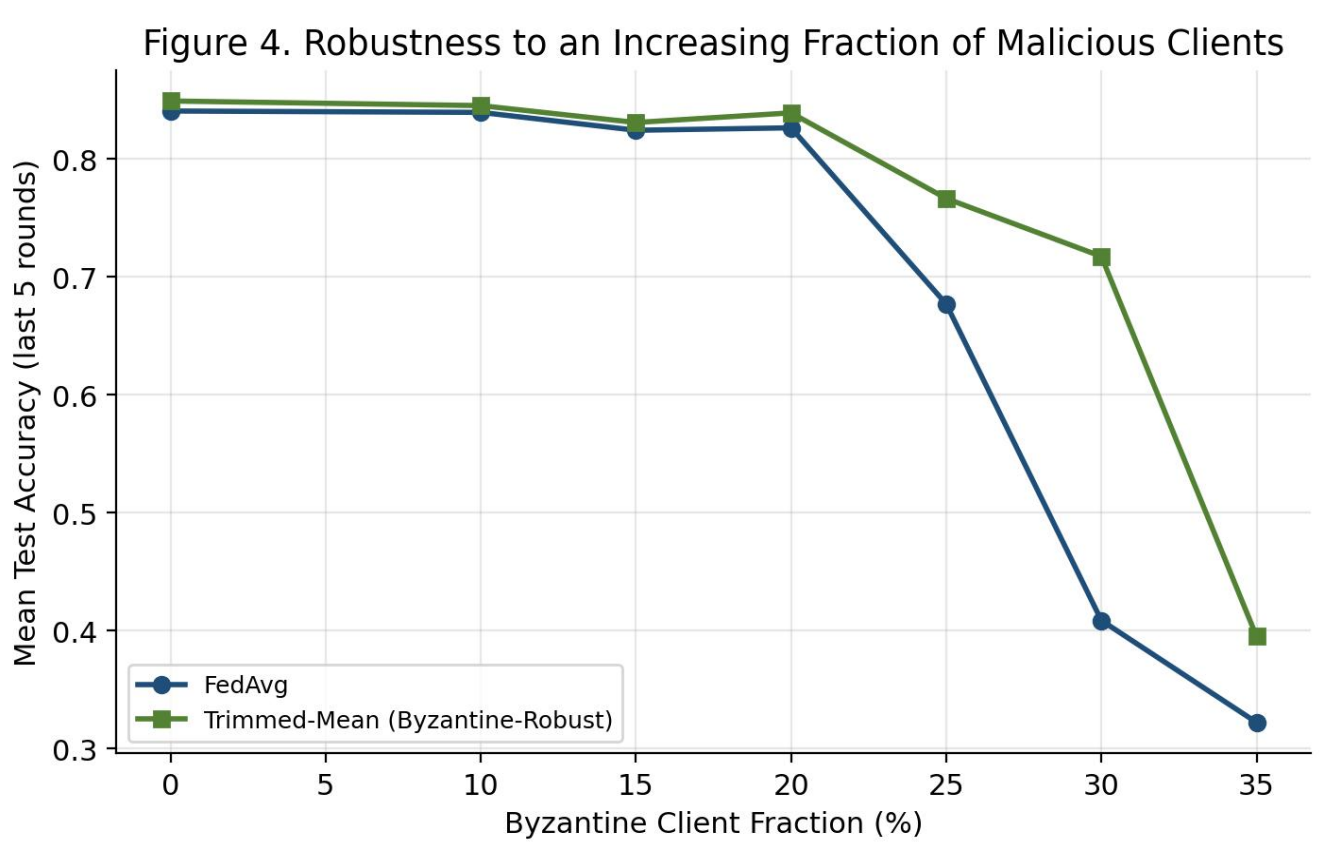

Fig. 5. Accuracy vs. Byzantine client fraction (0-35%).

## VI. SECURITY AND PRIVACY ANALYSIS

Under the threat model of Section III-A, DP-BR-FedAvg provides two guarantees that hold simultaneously. Against the honest-but-curious server, the client-level Gaussian mechanism (Eq. 4) combined with clipping (Eq. 1) gives an (epsilon, delta)-DP guarantee over the released sequence of global models, following the composability properties of Dwork and Roth [8]. Against Byzantine clients, the trimmed-mean rule (Eq. 2) bounds the influence of any subset smaller than beta, following Yin et al. [4], though this bound degrades as sigma increases since added noise widens the spread of honest updates.

Two residual risks are not addressed here. First, the framework assumes the server correctly executes the aggregation rule; an actively malicious server would require secure-aggregation or verifiable-computation techniques [5] as a complementary layer. Second, the guarantee is client-level, protecting the fact of an institution's participation, but does not by itself protect every possible record-level inference if an attacker accesses intermediate local checkpoints; deployments handling especially sensitive record-level data should consider record-level DP-SGD [2] within each client's local training in addition.

## VII. CONCLUSION

This paper presented DP-BR-FedAvg, combining client-level differential privacy with coordinate-wise trimmed-mean Byzantine-robust aggregation for cross-institutional deployments in banking and healthcare. On a simulated fraud/clinical-risk task with 20 clients, 25% Byzantine, the proposed framework roughly quadrupled minority-class F1-score relative to plain or privacy-only FedAvg, while a robust aggregator without privacy achieved the best raw performance among attacked configurations, quantifying the accuracy cost of the privacy layer. Privacy noise and robust aggregation interact: the robustness benefit narrows as noise increases, and the aggregator's own tolerance breaks down once the Byzantine fraction approaches the trim threshold. These findings argue against treating privacy and robustness as independently composable, and for co-tuning the noise multiplier and trim fraction against the anticipated threat levels in deployment.

## VIII. FUTURE RESEARCH DIRECTIONS

- Adaptive trim fraction and noise scheduling across rounds, based on observed update dispersion, to reduce the crossover effect in Section V-C.
- Tighter combined privacy-robustness accounting that formally reflects the sensitivity change introduced by trimmed-mean aggregation itself.
- Evaluation on real, de-identified institutional data under appropriate data-use and ethics agreements.
- Adaptive and colluding adversaries that adapt strategy across rounds or coordinate to evade trimmed-mean detection [17].
- Secure aggregation integration [5] so the server never observes individual client updates, closing the residual malicious-server risk.
- Regulatory and governance mapping of the (epsilon, delta) guarantee and Byzantine tolerance bound to specific banking/healthcare frameworks.
- Communication-efficient robust aggregation, building on [20], since trimmed-mean and Krum require full-precision updates each round.

## IX. REFERENCES

B. McMahan, E. Moore, D. Ramage, S. Hampson, and B. A. y Arcas, "Communication-efficient learning of deep networks from decentralized data," in Proc. 20th Int. Conf. Artificial Intelligence and Statistics (AISTATS), 2017, pp. 1273-1282.

M. Abadi, A. Chu, I. Goodfellow, H. B. McMahan, I. Mironov, K. Talwar, and L. Zhang, "Deep learning with differential privacy," in Proc. ACM SIGSAC Conf. Computer and Communications Security (CCS), 2016, pp. 308-318.

P. Blanchard, E. M. El Mhamdi, R. Guerraoui, and J. Stainer, "Machine learning with adversaries: Byzantine tolerant gradient descent," in Advances in Neural Information Processing Systems (NeurIPS), vol. 30, 2017, pp. 119-129.

D. Yin, Y. Chen, R. Kannan, and P. Bartlett, "Byzantine-robust distributed learning: Towards optimal statistical rates," in Proc. 35th Int. Conf. Machine Learning (ICML), 2018, pp. 5650-5659.

K. Bonawitz, V. Ivanov, B. Kreuter, A. Marcedone, H. B. McMahan, S. Patel, D. Ramage, A. Segal, and K. Seth, "Practical secure aggregation for privacy-preserving machine learning," in Proc. ACM SIGSAC Conf. Computer and Communications Security (CCS), 2017, pp. 1175-1191.

P. Kairouz et al., "Advances and open problems in federated learning," Foundations and Trends in Machine Learning, vol. 14, no. 1-2, pp. 1-210, 2021.

T. Li, A. K. Sahu, A. Talwalkar, and V. Smith, "Federated learning: Challenges, methods, and

future directions," IEEE Signal Processing Magazine, vol. 37, no. 3, pp. 50-60, 2020.

C. Dwork and A. Roth, "The algorithmic foundations of differential privacy," Foundations and Trends in Theoretical Computer Science, vol. 9, no. 3-4, pp. 211-407, 2014.

S. Truex, N. Baracaldo, A. Anwar, T. Steinke, H. Ludwig, R. Zhang, and Y. Zhou, "A hybrid approach to privacy-preserving federated learning," in Proc. 12th ACM Workshop on Artificial Intelligence and Security, 2019, pp. 1-11.

L. Zhu, Z. Liu, and S. Han, "Deep leakage from gradients," in Advances in Neural Information Processing Systems (NeurIPS), vol. 32, 2019, pp. 14774-14784.

M. J. Sheller et al., "Federated learning in medicine: facilitating multi-institutional collaborations without sharing patient data," Scientific Reports, vol. 10, article 12598, 2020.

N. Rieke et al., "The future of digital health with federated learning," npj Digital Medicine, vol. 3, article 119, 2020.

D. Byrd and A. Polychroniadou, "Differentially private secure multi-party computation for federated learning in financial applications," in Proc. 1st ACM Int. Conf. AI in Finance (ICAIF), 2020, pp. 1-9.

Q. Yang, Y. Liu, T. Chen, and Y. Tong, "Federated machine learning: Concept and applications," ACM Transactions on Intelligent Systems and Technology, vol. 10, no. 2, pp. 1-19, 2019.

R. C. Geyer, T. Klein, and M. Nabi, "Differentially private federated learning: A client level perspective," arXiv preprint arXiv:1712.07557, 2017.

C. Dwork, "Differential privacy," in Automata, Languages and Programming (ICALP 2006), Lecture Notes in Computer Science, vol. 4052, Springer, 2006, pp. 1-12.

E. Bagdasaryan, A. Veit, Y. Hua, D. Estrin, and V. Shmatikov, "How to backdoor federated learning," in Proc. 23rd Int. Conf. Artificial Intelligence and Statistics (AISTATS), 2020, pp. 2938-2948.

Z. Yang, M. Chen, W. Saad, C. S. Hong, and M. Shikh-Bahaei, "Energy efficient federated learning over wireless communication networks," IEEE Transactions on Wireless Communications, vol. 20, no. 3, pp. 1935-1949, 2021.

X. Cao, M. Fang, J. Liu, and N. Z. Gong, "FLTrust: Byzantine-robust federated learning via trust bootstrapping," in Proc. Network and Distributed System Security Symposium (NDSS), 2021.

J. Konecny, H. B. McMahan, F. X. Yu, P. Richtarik, A. T. Suresh, and D. Bacon, "Federated learning: Strategies for improving communication efficiency," arXiv preprint arXiv:1610.05492, 2016.

V. Mothukuri, R. M. Parizi, S. Pouriyeh, Y. Huang, A. Dehghantanha, and G. Srivastava, "A survey on security and privacy of federated learning," Future Generation Computer Systems, vol. 115, pp. 619-640, 2021.

L. Lyu, H. Yu, and Q. Yang, "Threats to federated learning: A survey," arXiv preprint arXiv:2003.02133, 2020.